\documentclass[12pt]{article}
\usepackage{times}
\usepackage{graphicx}
\usepackage{setspace,amssymb,bm}
\newenvironment{sciabstract}{\begin{quote} \bf}{\end{quote}}

\newcounter{lastnote}
\newenvironment{scilastnote}{%
\setcounter{lastnote}{\value{enumiv}}%
\addtocounter{lastnote}{+1}%
\begin{list}%
{\arabic{lastnote}.}
{\setlength{\leftmargin}{.22in}}
{\setlength{\labelsep}{.5em}}}
{\end{list}}

\title{Critical fluctuations and random-anisotropy glass transition in nematic elastomers}

\author{G.~Feio,$^{1,2}$ J. L. Figueirinhas,$^{1,3}$ A. R. Tajbakhsh,$^{4}$ E.M.~Terentjev$^{4}$\\
\\
\normalsize{$^{1}$CFMC, Universidade de Lisboa, Av. Prof. Gama Pinto 2,}\\
\normalsize{1649-003 Lisboa, Portugal}\\
\normalsize{$^{2}$Dept. Ci$\hat{e}$ncia dos Materiais, CENIMAT/I3N,
Fac. de Ci$\hat{e}$ncias e Tecnologia}\\
\normalsize{Universidade Nova de Lisboa, Caparica, P-2829-516 Caparica, Portugal}\\
\normalsize{$^{3}$IST, Av. Rovisco Pais 1049-001 Lisboa, Portugal}\\
\normalsize{$^{4}$Cavendish Laboratory, University of Cambridge, J. J. Thomson Avenue,}\\
\normalsize{ Cambridge CB3 OHE, U.K. }\\
}

\date{}
\begin{document}

\baselineskip24pt
\maketitle

\begin{sciabstract}
We carry out a detailed deuterium NMR study of local nematic
ordering in polydomain nematic elastomers. This system has a close
analogy to the random-anisotropy spin glass. We find that, in
spite of the quadrupolar nematic symmetry in 3-dimensions
requiring a first-order transition, the order parameter in the
quenched ``nematic glass'' emerges via a continuous phase
transition. In addition, by a careful analysis of the NMR line
shape, we deduce that the local director fluctuations grow in a
critical manner around the transition point. This could be the
experimental evidence for the Aizenman-Wehr theorem about the
quenched impurities changing the order of discontinuous
transition.
\end{sciabstract}

\newpage

\section*{Introduction}
The debate over the nature of the spin-glass phases in finite
dimensions (and in experiment) is still vibrant in the modern
literature \cite{moore1,katz,bouchaud}. The key question is about
the nature of the symmetry that is broken by the spin-glass
transition. The simplest idea is that the broken symmetry is the
standard up-down, Ising symmetry. Even a random-anisotropy
Heisenberg system apparently maps into the Ising spin glass
\cite{moore2}. Low energy excitations of arbitrarily large size
confer critical properties to the system; in particular the
non-linear susceptibility diverges in the whole spin-glass phase.
However, because the up-down symmetry is sensitive to the external
magnetic field, there cannot be a strict phase transition in
non-zero field, only a dynamical crossover. In mean-field, or
continuum models, on the other hand the broken symmetry is in fact
the Parisi `replica symmetry' \cite{parisi}, which encodes
mathematically the presence of many possible ordered phases,
unrelated to one another by simple transformations. The external
field cannot select one out of a very large number of possible
frozen configurations. Therefore in mean-field one expects, and
indeed finds, a true phase transition even in non zero field --
the celebrated de Almeida-Thouless (AT) line \cite{almeida}.

There has been relatively little study on how the quenched
disorder influences the systems whose pure versions undergo a
first-order phase transition. This question was first addressed by
Imry and Wortis \cite{wortis} who showed that inhomogeneities may
cause local variations of the transition temperature inside the
sample. Provided that the cost in interface energy is not great,
bubbles of the `wrong' phase are formed, eventually leading to a
substantial rounding of the transition. A theorem due to Aizenman
and Wehr \cite{aizenman} shows that in less than two dimensions in
a system with quenched random impurities there can be no phase
coexistence at the transition, and therefore no latent heat.
Therefore these systems are expected to always exhibit a
continuous transition. The influence of quenched impurities
coupling to the local energy density has been extensively studied
by Cardy \cite{cardy}. It was found that, depending on the
specific values of parameters of the pure system, such as the
latent heat and the surface tension, the disorder-affected phase
transition is fluctuation-driven and can either be first or second
order. In both cases of spin-glass and quenched first-order
systems the experimental measurement of equilibrium transition
characteristics is very difficult and only few indirect results
are available; most of the studies, such as neutron spin-echo
experiments \cite{mezei} only access dynamic quantities and
correlation functions.

A physical system, in which the sources of quenched orientational
disorder can be coarse-grained to be represented by a weak random
field, is liquid crystalline elastomers \cite{emt1}. Quenched
disorder is intrinsically present in all elastomers as a direct
result of their synthesis. In the simplest situation, the network
crosslinking takes place in the isotropic phase, in which case the
local anisotropy axis of each crosslinking group is randomly
oriented \cite{fridrikh}. Once the polymer network is formed, the
configuration of the crosslinks remains quenched, and the
underlying nematic \cite{petridis} or smectic \cite{olmsted} phase
transition takes place with their effect in the background. Other
liquid crystalline systems with random disorder include the
nematic in pores of silica gels \cite{bellini}, polymer-stabilized
and polymer-dispersed liquid crystals, e.g. \cite{pdlc}. The
source of the random disorder in such cases is the surface
anchoring of nematic director on the walls of the random porous
matrix on a length scale similar, or greater than the
characteristic period of resulting director textures. In nematic
elastomers the random disorder arises from defects in the polymer
network structure and cross-links quenched during the synthesis,
on a much smaller length scale, which allows for coarse-graining
and the continuum mean-field description.

In this work we carry out the deuterium NMR experiments to measure
the local nematic order parameter in the `polydomain' nematic
elastomer system, that is, the equilibrium symmetry broken state
with the orientational texture fully analogous to the spin glass.
Strictly, the nematic order parameter is a tensor with quadrupolar
symmetry, arising from the averaging of axes ($\bm{\hat{u}}$) of
rod-like molecular groups: $Q_{ij}=\langle \hat{u}_i \hat{u}_j -
\frac{1}{3} \delta_{ij} \rangle = S(T) [n_i n_j - \frac{1}{3}
\delta_{ij}]$ in three dimensions. In the system with quenched
disorder one must distinguish between the time- and the ensemble
averaging, since the principal axis $\bm{n}$ of this order
parameter (called the nematic director) randomly varies in space.
NMR is the only technique we know that can access the local value
of the scalar order parameter $S(T)$ in spite of the powder
averaging over the $\bm{n}(\bm{x})$ orientations.

There have been many measurements of the nematic order parameter
in aligned nematic elastomers, in which case a variety of
experimental techniques are available. The alignment in these
materials can be installed in two ways: (1) by applying a uniaxial
field (usually mechanical stress) to an original polydomain
texture and passing through the polydomain-monodomain transition
\cite{fridrikh,emt2}, or (2) by crosslinking the network in a
state uniformly aligned by a mechanical, electric, or surface
field \cite{kupfer,urayama}. In both cases, of course, one does
not expect a true phase transition and indeed all the reports find
the varying degrees of supercritical crossover, see
\cite{lebar,cordoyanis} and references therein.

In contrast, we deliberately devise the physical system as close
as possible to the equilibrium `nematic glass' state. Apart from
crosslinking the network in a fully isotropic state, carefully
avoiding any stray aligning influences, we also choose the type of
crosslinking that is creating as weak as possible orientational
effect (cf. Fig.~\ref{chem}). There have been many reports showing
that using a more robust rod-like crosslinking group results in
much stronger orientational effect and possibly undermines the
approximations required for the continuum coarse-graining of the
random-anisotropy field. It would be very interesting to conduct
comparative studies, with varying the characteristics of random
field by changing crosslinking density and molecular structure.
However, in this first study we concentrate on the system where we
are safely within the limits of weak random field, and the
resulting `nematic glass' phase. There are two key results we
report here. First of all, after careful analysis of the D-NMR
signal, we are able to obtain the temperature dependence of the
equilibrium local order parameter $S(T)$, which has a clear
critical behavior $S \propto |T-T_c|^{0.22}$ according to our best
fit. This was the principal aim of our study. However, in
addition, we found that the NMR line width (which reflects the
loss of local spin mobility) has a very clear $\lambda$-shaped
peak at the transition point $T_c$. We associate this peak with
the critical growth of correlation length of fluctuations on both
sides of the transition, which further supports the conclusion
about the continuum transition from the isotropic into the
`nematic glass' phase.

\section*{Deuterium NMR analysis}

The spin Hamiltonian for deuterium in a high magnetic field
$\bm{B}_{0}$ is dominated by the Zeeman and quadrupolar
interactions \cite{abragam} with the latter amenable to a first
order perturbation theory calculation in order studies. In the
absence of motion a carbon bound deuterium gives rise to a doublet
absorption spectra centered at the Larmor precession frequency and
with a splitting given by
\begin{equation}
w =\frac{6\pi}{2} \nu_{Q} \left( \frac{3}{2} \cos^2 \theta -
\frac{1}{2}+
  \frac{1}{2}\eta \sin^2 \theta \cos 2\phi \right)
  \label{larmor}
\end{equation}
where $\theta$ and $\phi$ are the polar and azimuthal angles that
define the orientation of the external magnetic field in the
principal frame of the electric field gradient (EFG) tensor
$V_{ij}$ at the nucleus site. The characteristic frequency scale
$\nu_{Q}=(eQ/h) V_{zz}$ is called the quadrupolar coupling
constant which for s$p^{2}$ bonds is $\approx$185~kHz
\cite{natoveracini} ($e$ is the electron charge, $Q$ --
quadrupolar moment of ${}^2$H nucleus, and $h$ the  Planck
constant). $\eta=(V_{xx}-V_{yy})/V_{zz}$ is the biaxial asymmetry
parameter, which is small and negligible here.
 In a polydomain sample the contributions
from all $\theta$, $\phi$ orientations generate a characteristic
spectral pattern that for a 3D powder average is known as a `Pake
pattern' \cite{pake} which allows a direct measurement of the
time-averaged $\overline{\nu}_{Q}$. The corresponding EFG tensor
component is directly related to the orientational nematic order
parameter via $\overline{V}_{zz} = v^m  S$, where $v^m$ is a
constant for a given molecule \cite{natozanoni}. The Pake spectra
were fitted to the doublet expression:
 \begin{equation}
G(\omega) =\int_{\Omega} \left[ h(\omega - \textstyle{\frac{1}{2}}
{w(\theta,\phi, \overline{\nu}_{Q})}, \Delta \omega,\alpha)+
h(\omega + \textstyle{\frac{1}{2}}
{w(\theta,\phi,\overline{\nu}_{Q})},
 \Delta \omega, \alpha) \right] \,  \sin\theta d\theta d\phi
 \label{pake}
 \end{equation}
using a conjugate directions method \cite{numericalrecipes}, where
$w(\theta,\phi,\overline{\nu}_{Q})$ is given by (\ref{larmor}) in
the dynamic case, the splitting amplitude $\overline{\nu}_{Q}$ and
the individual line width $\Delta \omega$ are fitting parameters,
and $h( \omega, \Delta \omega,\alpha )$ is the specific line shape
function.

\begin{figure} %[b]
  \begin{center}
\centering \resizebox{0.5\textwidth}{!}{\includegraphics{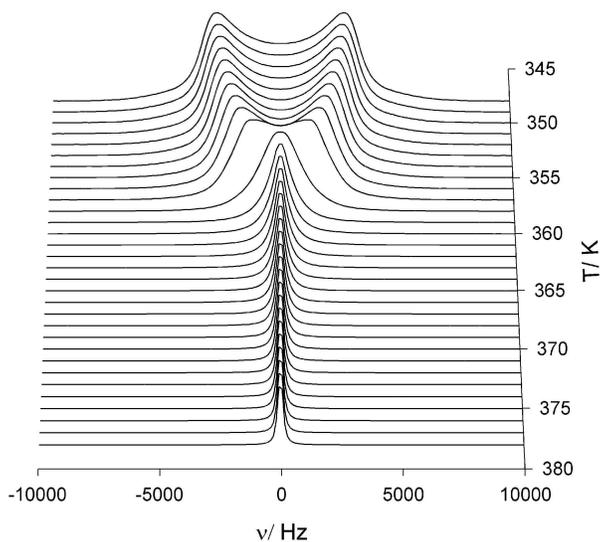}}
\caption{ \textbf{D-NMR spectra on changing temperature}.
Composite representation of Pake spectra evolution on cooling the
system from the isotropic state. Note the absence of any
coexistence between the emerging nematic splitting and the
isotropic central peak. } \label{spectra3D}
  \end{center}
\end{figure}

The doublet and powder spectra discussed above are formed by
single and multiple pairs of structured lines, respectively,
determined by the transverse relaxation occurring in the system
\cite{kimmich}. When the correlation time $\tau_{c}$ of the
fluctuating quadrupolar interaction is small,
 $\tau_{c}\nu_{Q} \ll  1/(2\pi)$ (the fast-motion limit),
the line structure acquires a Lorentzian form corresponding to an
exponentially decaying transverse magnetization in the rotating
frame, $e^{-t/T_{2}^{*}}$. Here $T_{2}^{*}$ is the  decay time of
transverse magnetization related to the frequency width at half
height $\Delta \omega$ of the Lorentzian line by
$T_{2}^{*}=2/\Delta \omega$:
 \begin{equation} h(\omega) = {\rm
FT}\left\{ e^{- t/T_{2}^{*}}  \right\} \equiv
 \frac{T_{2}^{*}}{1 + (T_{2}^{*} \omega)^2} . \label{lorentz}
 \end{equation}
As the molecular motion slows down, the coherence between the
nuclear spins reduces faster, making the spins precess at different
frequencies (larger $\Delta \omega$); in real time the correlation
between the signals from the different spins reduces and the NMR
signal decays faster (smaller $T_{2}^{*}$). In the opposite, slow-motion
limit, $\tau_{c}\nu_{Q} \gg 1/(2\pi)$, a Gaussian shape becomes a
good approximation for the line structure \cite{abragam}, given by
the Fourier transform of
 \begin{equation}
h(\omega) = {\rm FT}\left\{ e^{-(t/\tau_{g})^2} \right\} \equiv
 \sqrt{\pi} \, \tau_{g} \, e^{- \omega^2 \tau_{g}^2/4}
 \label{gauss}
 \end{equation}
where the signal decay time $\tau_{g}$ is related to the frequency
width by $\tau_{g} = 4\sqrt{\ln 2}/\Delta \omega$. Figure
\ref{spectra3D} gives the composite representation of a sequence
of D-NMR scans on cooling from the isotropic state, illustrating
the widening of the central peak, and then the emergence of
characteristic nematic splitting. It is very important to note the
complete absence of any coexistence, which is a necessary feature
of a first-order transition and which has been observed in
previous NMR experiments on nematic elastomers
\cite{lebar,cordoyanis} (which, for various reasons, always had a
local stress frozen into the network preventing critical
fluctuations).

It was suggested by Brereton and others
\cite{brereton1,brereton2,warner,deloche} that in isotropic polymer networks, even
in the absence of an ordering field, the averaging of second-rank
tensorial interactions from nuclear spins is not complete due to the local
anisotropy imposed by crosslink constraints. Nematic interactions
between chain segments have also been shown to contribute to the
widening of spectra. This gives rise to the prediction of a
characteristic absorption spectra with a cusp at zero frequency
\cite{brereton1,warner,deloche}. In our case, the deuterium labels are
placed in the mesogenic units, away from the backbone and these
constraints are expected to have a small effect on the NMR line
shape, contributing at most to increase the line width in a weakly temperature
dependent mode. Indeed the high-temperature spectra closely follow the
classical Lorentzian shape.

\begin{figure} %[b]
  \begin{center}
\centering \resizebox{0.75\textwidth}{!}{\includegraphics{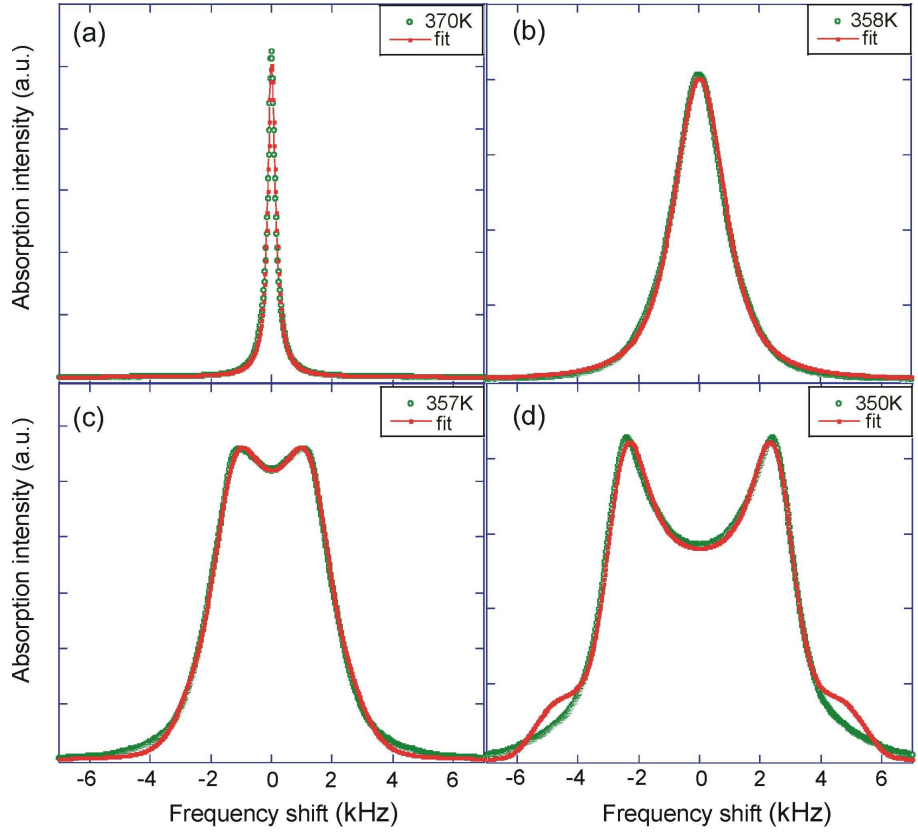}}
\caption{ \textbf{Fitting D-NMR spectra}. Plots show the raw data
for D-NMR spectra at several key temperatures (labelled on plots),
above and below the phase transition point. The fits are obtained
by the combined Lorentzian-Gaussian model discussed in the text.
Later in the paper we shall find the nematic transition point $T_c
= 358.01$K, so it is important to see no splitting in (b) and a
pronounced splitting in (c), with no co-existence effects as in
\cite{lebar,cordoyanis}. } \label{spectra}
  \end{center}
\end{figure}

Figure~\ref{spectra} gives the more detailed absorption spectra
obtained in the D-NMR experiment in characteristic regimes.  The
high-temperature isotropic phase has a narrow line that closely
matches the Lorentzian form, Fig.~\ref{spectra}(a),  as expected
in a classical process of fast rotational diffusion. In the
nematic phase molecular motion slows down and collective modes set
in and the orientational relaxation time of mesogenic groups
attached to the polymer backbone increases significantly. In a low
molecular weight nematic this time is typically of the order of
several times $10^{-7}$s, which is very much shorter than the
characteristic time of NMR. Measurements of rotational viscosity
$\gamma_1$ \cite{stille1,stille2}, which is proportional to the
relaxation time \cite{osipov,chan}, show an increase of over 4
orders of magnitude in the nematic phase of elastomers, comparing
to the low-molecular weight nematics. This suggests that the
correlation time of rod-like groups attached to the backbone
becomes closer to the characteristic NMR time
$(2\pi\nu_{Q})^{-1}$, as soon as the nematic mean field sets in;
indeed we find that the Gaussian form is more appropriate to
describe the line structure in the nematic phase.

\begin{figure} %[b]
  \begin{center}
\centering \resizebox{0.5\textwidth}{!}{\includegraphics{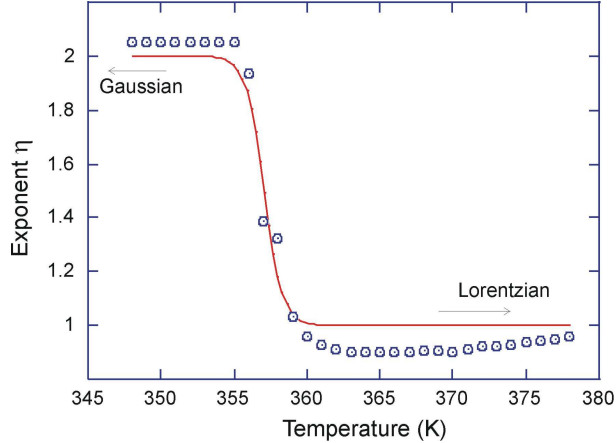}}
\caption{ \textbf{Lorentzian vs. Gaussian NMR line}. The values of
exponent $\alpha$ obtained from a three-parameter fit ($\circ$)
are plotted for each temperature. Clearly the values of $\alpha =
1$ and $2$ are approached far from the transition point. The solid
line is the best fit by (\ref{alpha}), which we then use as the
fixed value for $\alpha(T)$ function.
 } \label{fixexp}
  \end{center}
\end{figure}

In order to test fitting the NMR spectra in the whole temperature
range we have applied a model with the variable exponent
$\alpha(T)$, numerically Fourier-transforming $\exp \left[ -
\pi\,( t \, \Delta f)^\alpha \right]$. Figure~\ref{fixexp} shows
the results for the fitted empirical exponent $\alpha(T)$. It is
clear that the basic expectation of the crossover between the
Lorentzian regime in the isotropic phase ($\alpha \approx 1$ at
$T>T_c$) crosses over to the Gaussian regime in the nematic phase
at $T<T_c$, where $\alpha \approx 2$. To reduce the unnecessary
freedom of multi-parameter fitting, we choose to assign a fixed
form to the variable exponent $\alpha(T)$ so that it describes the
fundamental crossover between $\alpha=1$ and $2$, as shown in the
plot:
\begin{equation}
\alpha (T) = 1.5 - 0.5 \, \tanh \left[ \frac{T-T^*}{m} \right] .
\label{alpha}
\end{equation}
The closest fit is achieved with $T^*= 357.1$K and the width of
temperature crossover $m= 1.62$K. Note that (as we shall see
below) the actual critical point of the transition is about a
degree higher, $T_c = 358$K. This shift is natural, since the
crossover from the diffusive to the slow-relaxation regime
(Lorentzian to Gaussian line shape) occurs when the local nematic
mean field grows sufficiently strong to slow down the molecular
rotation rate closing in the characteristic NMR frequency. With
this choice of $\alpha(T)$ we only have two fitting parameters,
$\overline{\nu}_{Q}$ and $\Delta \omega$, which are obtained as
functions of temperature. The quality of line fits is very good in
the relevant sections, as is evident from Fig.~\ref{spectra},
although the long tails of the peaks at very low temperatures
deviate slightly as illustrated in the (d) plot for $350$K.

\section*{Nematic-isotropic transition}

Having established the rules of analysis of our D-NMR signal, we
then proceed with fitting the Pake pattern (\ref{pake}) to all the
spectra. In this fitting we only allow two free parameters: the
average quadrupolar coupling constant, $\overline{\nu}_Q$, and the
characteristic width of the distribution, $\Delta \omega \equiv
2\pi \Delta f$. The first parameter gives directly the magnitude
of the nematic order $S(T)$, in this case defined as the
time-average of orientational motion of individual rod-like units.
In the non-ergodic system such as polydomain ``nematic glass'' it
is essential to distinguish between this, and the spatially
averaged quantities (such the outputs of X-ray, or optical
dischroism measurements). There is an effective averaging of the
output signal over the sample volume, but in the NMR experiment it
takes place after the field of the tensor order parameter
$Q_{ij}(\bm{r})$ is established at every point at each
temperature. In this sense, our analysis is based on the
assumption \cite{petridis} that the scalar value of $S$ is uniform
throughout the polydomain director texture in the sample. If this
was not the case, i.e. the magnitude of the nematic order varied
in space (e.g. across domain boundaries or in disclination cores),
then our clean double-peak Pake pattern would not be observed deep
in the nematic phase. This alone is a very important observation,
making a stark contrast with the classical Schlieren texture of a
disclination-coarsening nematic liquid \cite{yurke}.

\begin{figure} %[b]
  \begin{center}
\centering \resizebox{0.55\textwidth}{!}{\includegraphics{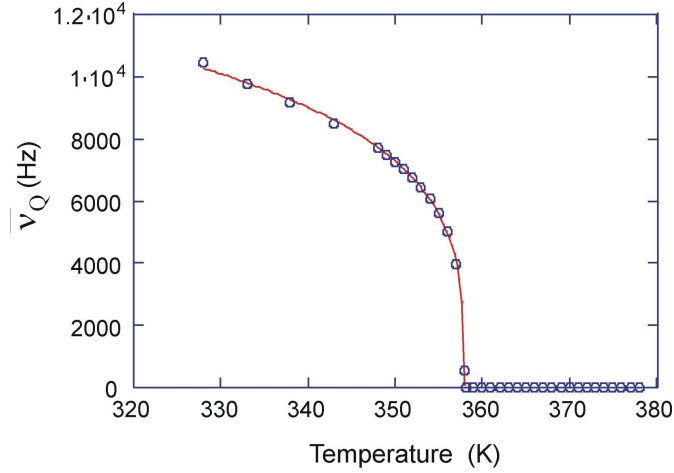}}
\caption{ \textbf{Nematic order parameter}. Plot of the average
quadrupolar coupling constant $\overline{\nu}_Q$, obtained from a
2-parameter fit of each spectrum, against temperature. Apart from
a constant dimensional factor, $\overline{\nu}_Q$ directly
represents the local nematic order parameter $S(T)$. }
\label{order}
  \end{center}
\end{figure}

Figure \ref{order} shows the fitted values of $\overline{\nu}_Q$
against temperature. It is clear, as it was visually apparent from
the shape of NMR peaks in figure~\ref{spectra3D}, that the nematic
order sets in in a critical fashion. The fit to the data suggests
the variation $S \propto |T-T_c|^{0.22}$, with the critical point
$T_c = 358.01$K. Importantly, the fitting gives values $S=0$ above
$T_c$ unambiguously, indicating that no supercritical or
``paranematic'' effects take place in our system. This is another
important point, contrasting with many other studies. It is
well-known that monodomain (macroscopically aligned, crosslinked
under uniaxial external field) nematic elastomers have a diffuse
supercritical transition, with its sharpness decreasing with more
rigid crosslinkers or higher external fields imposed at formation.
The recent detailed calorimetric study \cite{cordoyanis} gives the
details of this, and the earlier literature. There are very few
studies of polydomain systems, and none that examine the soft
un-entangled networks (obtain by de-swelling, with small flexible
crosslinks). Due to our preparation, the system we study here is
close to true criticality, which is a remarkable effect of
quenched random-anisotropy disorder considering that the
underlying quadrupolar nematic ordering is a first-order
transition.

\begin{figure} %[b]
  \begin{center}
\centering \resizebox{0.95\textwidth}{!}{\includegraphics{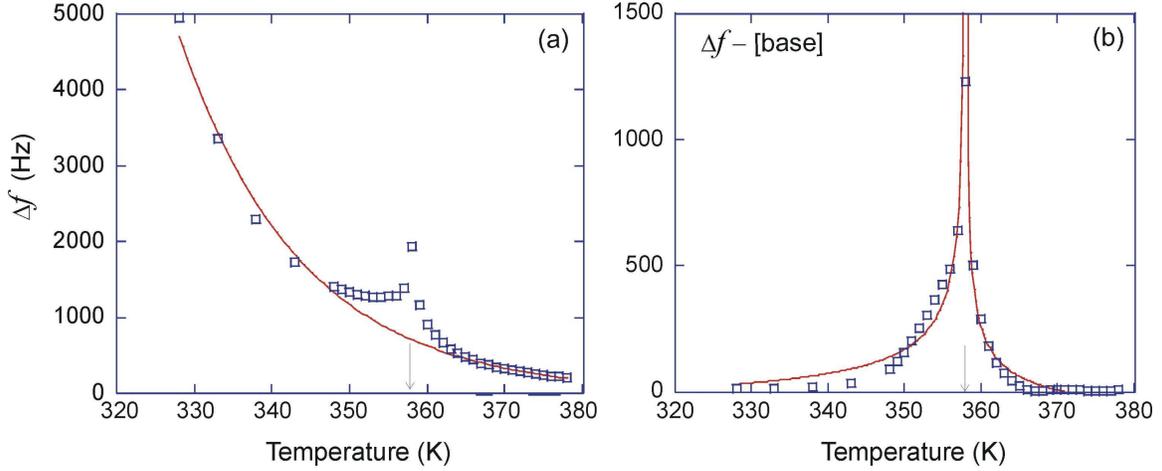}}
\caption{ \textbf{Critical fluctuations}. (a) Plot of the
frequency width $\Delta f \equiv \Delta \omega / 2\pi$, obtained
from a 2-parameter fit of each spectrum, against temperature. The
solid line is a baseline fit for a polymer network that gradually
approaches its glass transition, resulting in reduced segment
mobility and the associated widening of the spectra. \ (b) Plot of
the frequency width with the baseline subtracted, illustrating the
particular widening during the transition. The solid line on both
sides of $T_c$ is a fit by $|T-T_c|^{-0.5}$, as in Landau-Ginzburg
model of critical divergence of fluctuation correlation length. }
\label{fluct}
  \end{center}
\end{figure}

The second evidence for the critical transition is found in the
behavior of our second fitting parameter: the distribution width
$\Delta \omega$, which is proportional to the correlation time of
orientational motion of mesogenic groups. Figure \ref{fluct}(a)
shows the increase of $\Delta \omega$ on cooling the elastomer
network. There are clearly two separate effects: the overall
increase representing the slowing down of molecular motion on
approaching the structural glass transition in a polymer network,
and the pronounced peak around the phase transition. The solid
line in figure \ref{fluct}(a) shows the exponential fit of the
baseline data to the Vogel-Fulcher exponential $\Delta \omega
\propto \exp \left[ A/(T-T_\beta) \right]$, with $A = 3200$ and
$T_\beta=120$K. It is not unexpected to find the cutoff
temperature $T_\beta$ so much lower than the ``ordinary'' polymer
glass transition $T_g \approx 300$K in our material: the latter
represents the so-called $\alpha$-relaxation, or freezing of the
polymer backbone motion. Our NMR experiment probes the motion of
side-groups, that is $\beta$- or even $\gamma$-relaxation, which
freezes at much lower temperatures. Importantly, this baseline
increase is continuous across both the Lorentzian and the Gaussian
regimes on both sides of the nematic transition.

Figure \ref{fluct}(b) plots the $\Delta \omega$ peak around the
transition, with the baseline polymer dynamics taken out. The
solid lines on both sides of $T_c$ are fits to the critical
$|T-T_c|^{-b}$, using the value $T_c=358.01$K from the analysis in
figure \ref{order} and returning the same exponent $b = 0.5$ on
both sides of $T_c$. We suggest that this behavior represents the
growth of correlations of critical fluctuations near the
transition point. As the size of correlated nematic fluctuations
increases near the critical point, $\xi \propto |T-T_c|^{-\nu}$,
the time for (collective) re-orientation of rod-like units within
this volume increases proportionally. Curiously, the exponent we
obtain from the best fit is very close to the mean-field
$\nu=1/2$, which connects with the earlier discussion of the
mean-field approximation and the de Almeida-Thouless line.

\section*{Summary}

A detailed study of deuterium NMR in polydomain nematic elastomers
suggests that the structure of this state is analogous to
random-anisotropy spin glass. Because of this analogy, the results
for the local symmetry breaking, order parameter and fluctuations
near the phase transition have a broad significance for the whole
field. The local magnitude nematic order parameter is homogenous
throughout the system in spite of the director correlations being
short-range in the replica-symmetric approximation. Analysis of
NMR lines at different temperatures suggests that the transition
from the isotropic high-temperature phase is of continuous
critical nature. Since this transition (quadrupolar order with
weak quenched random-anisotropy field) has not been studied with
renormalization group, we cannot assign any particular
significance to the apparent critical exponents for the order
parameter magnitude, $S \propto |T-T_c|^{0.22}$ and correlation
length of critical fluctuations, $\xi \propto |T-T_c|^{-0.5}$.
However, the fact that we see an apparent critical behavior in a
system that undergoes a first-order transition in its pure state
is of remarkable importance. This could be the first experimental
study that confirms the Aizenman-Wehr theorem predicting such an
effect. Secondary findings of this work, on the continuous change
between the Lorentzian and Gaussian NMR line shape and on the
continuous slowing down of $\beta$-relaxation of polymer
side-groups on cooling the elastomer, are also of great interest
in their respective fields.

\section*{Methods}

\begin{figure} %[b]
  \begin{center}
\centering \resizebox{0.65\textwidth}{!}{\includegraphics{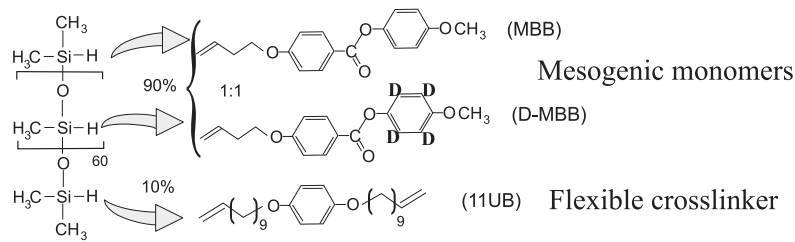}}
\caption{ \textbf{Composition of nematic elastomer}. The classical
polysiloxane nematic elastomer is obtain by attaching the rod-like
mesogenic units (MBB) to the flexible siloxane chain. 50\% of
these units were deuterated (D-MBB), symmetrically to preserve the
molecular symmetry as shown in the sketch. Permanent network
crosslinking has been achieved by reacting with 2-functional
flexible molecule 11UB, which attaches to two backbone chains. }
\label{chem}
  \end{center}
\end{figure}

\noindent \textbf{Preparation}. \  All starting materials and
samples of side chain siloxane liquid crystalline elastomers were
prepared in the Cavendish Laboratory following the hydrosilation
procedures developed by H. Finkelmann \cite{siloxane}. The polymer
backbone was a poly-methylhydrosiloxane with approximately 60 Si-H
units per chain, obtained from ACROS Chemicals. The pendant
mesogenic groups were 4'-methoxyphenyl-4-(1-buteneoxy) benzoate
(MBB), as illustrated in figure~\ref{chem}, attached to the
backbone via the reaction of of Si-H group with the vinyl bond.
For good quality deuterium NMR signal, we used 50\% of the MBB
units with the fully deuterated outer benzene ring.  All networks
were chemically crosslinked via the same reaction, in the presence
of commercial platinum catalyst COD, obtained from Wacker Chemie,
with the flexible di-functional crosslinking group 1,4
di(11-undeceneoxy)benzene (11UB) also shown in Fig.~\ref{chem}.
The crosslinking density was calculated to be 10~mol\% of the
reacting bonds in the siloxane backbone, so that on average each
chain has 9 mesogenic groups between crosslinking sites.

Our aim has been to obtain a true polydomain state of the nematic
phase, which can only be generated when the crosslinking is fully
random and no internal stresses of entanglements are frozen into
the resulting network. For this, we have completed the
crosslinking reaction in the highly swollen (in toluene) isotropic
phase of the polymer. After the completion of all reactions, the
samples were slowly de-swollen so that the dry elastomer network
did not end up over-entangled.

\noindent \textbf{Testing procedure}. \

The NMR data was collected using a Bruker Avance II+ 300 MHz
spectrometer working at a resonance frequency for deuterium of
46.1 MHz. The solid echo sequence with a pulse separation of
20$\mu$s and a $\pi$/2 pulse width of 5$\mu$s was used to obtain
the NMR spectra in a high power, wide line, probe head. The
elastomer sample consisting of a thin stripe of material was
placed over holding glass rods evenly placed horizontally within a
cylindrical teflon case. The measurements where performed after
having heated the sample up to 378K (more than 15K above the
nematic-isotropic transition) at a rate of 5K/min. After a
stabilization period of 30 min the first measurement of the
highest-temperature state was taken. Upon conclusion, the sample
was slowly cooled, at a rate of 1K/min, to the second temperature
to be analyzed, followed by an equilibration time delay of 10 min
at this temperature. Consecutive NMR measurements were obtained on
decreasing the temperature from 378K to 348K, at 1K intervals, all
preceded by the identical equilibration procedure. At lower
temperatures, down to 328K, further measurements were recorded 5K
apart. Four thousand scans (with a recycle delay of 500ms) were
collected in each measurement to reach acceptable signal to noise
ratios.

%\bibliography{scibib}
\bibliographystyle{Science}

\begin{scilastnote}
\item The authors wish to thank M.A. Moore, C. Zannoni, B. Zalar, D. Finotello,
M.H. Godinho and
L. Petridis for many useful discussions. This work was carried out
with the support of the EPSRC and Portuguese Science and
Technology Foundation (FCT) through projects: POCI/CTM/56382/2004,
POCI/CTM/61293/2004 and PTDC/FIS/65037/2006.
\end{scilastnote}

\end{document}